\documentclass[twocolumn,superscriptaddress,showpacs]{revtex4-1}
\usepackage{graphicx}  
\usepackage{dcolumn}   
\usepackage{bm}        
\usepackage{amssymb}   
\usepackage{amsfonts}
\usepackage{amsmath}
\usepackage{hyperref}

\newcommand{\E}{\operatornamewithlimits{E}}

\begin{document}
\title{Bounds of memory strength for power-law series}

\author{Fangjian Guo}
 \affiliation{CompleX Lab, Web Sciences Center, University of Electronic Science and Technology of China, Chengdu 611731, People's Republic of China}
 \affiliation{Department of Computer Science, Duke University, Durham, NC 27708, USA}
\author{Dan Yang}
 \affiliation{CompleX Lab, Web Sciences Center, University of Electronic Science and Technology of China, Chengdu 611731, People's Republic of China}
 \affiliation{Big Data Research Center, University of Electronic Science and Technology of China, Chengdu 611731, People's Republic of China}
\author{Zimo Yang}
 \affiliation{CompleX Lab, Web Sciences Center, University of Electronic Science and Technology of China, Chengdu 611731, People's Republic of China}
\author{Zhi-Dan Zhao}
 \affiliation{CompleX Lab, Web Sciences Center, University of Electronic Science and Technology of China, Chengdu 611731, People's Republic of China}
\author{Tao Zhou}
 \email{zhutou@ustc.edu}
 \affiliation{CompleX Lab, Web Sciences Center, University of Electronic Science and Technology of China, Chengdu 611731, People's Republic of China}
 \affiliation{Big Data Research Center, University of Electronic Science and Technology of China, Chengdu 611731, People's Republic of China}

\begin{abstract}
Many time series produced by complex systems are empirically found to follow power-law distributions with different exponents $\alpha$. By permuting the independently drawn samples from a power-law distribution, we present non-trivial bounds on the memory strength (1st-order autocorrelation) as a function of $\alpha$, which are markedly different from the ordinary $\pm 1$  bounds for Gaussian or uniform distributions. When $1 < \alpha \leq 3$, as $\alpha$ grows bigger, the upper bound increases from 0 to +1 while the lower bound remains 0; when $\alpha > 3$, the upper bound remains +1 while the lower bound descends below 0. Theoretical bounds agree well with numerical simulations. Based on the posts on Twitter,  ratings of MovieLens, calling records of the mobile operator Orange, and browsing behavior of Taobao, we find that empirical power-law distributed data produced by human activities obey such constraints. The present findings explain some observed constraints in bursty time series and scale-free networks, and challenge the validity of measures like autocorrelation and assortativity coefficient in heterogeneous systems.
\end{abstract}

\maketitle

\section{Introduction}

Heterogeneity and memory strength are two remarkable features in characterizing time series produced by complex systems. Heterogeneity quantifies the extent that the distribution of time series differs from a normal distribution. In particular, many real time series, denoted by $\{t_1,t_2,\cdots\}$, are empirically found to follow power-law distributions $p(t)\sim t^{-\alpha}$ \cite{Clauset2009}. Examples include the magnitudes of earthquake \cite{Pisarenko2003}, the intensity of wars \cite{Roberts1998}, the severity of terrorist attacks \cite{Clauset2007}, the strength of fluctuations in finance market \cite{Wang2001}, the time intervals between human-activated events \cite{Barabasi2005}, and so on. Notice that, when talking about time intervals, $t_i$ stands for the time interval between the occurrences of the $i$th and $(i+1)$th events, but not the exact time the $i$th event happens. Memory strength characterizes the correlation between neighboring or nearby elements: a positive memory strength indicates that a high-value (low-value) element is likely to be followed by some high-value (low-value) elements. Memory strength effects are observed in a large variety of real-world time series \cite{Palma2007}, such as natural phenomena (e.g., earthquake and daily precipitation) \cite{Goh2008}, online human activities (e.g., visits on movies, books and music) \cite{Zhao2012}, physiological signals (e.g., respiratory and liver cirrhosis time series) \cite{Shirazi2013}, and so on. Such memory strength effects also play a significant role in modeling spatial mobility \cite{Szell2012,Choi2012,Zhao2013} and temporal activities \cite{Vazquez2007,Han2008} of human beings.

Here we adopt the simplest measure, first-order autocorrelation \cite{Goh2008}, to characterize the memory strength effect of a finite time series $\{t_1,t_2,\cdots,t_n\}$, namely the Pearson's correlation coefficient between $ \{t_{1}, t_{2}, \cdots, t_{n-1} \}$ and $ \{t_{2}, t_{3}, \cdots, t_{n}\} $, defined as
\begin{equation}
	M = \frac{1}{n-1} \sum_{i=1}^{n-1} \frac{(t_i - m_1)(t_{i+1} - m_2)}{\sigma_1 \sigma_2},
\end{equation}
where $m_1, m_2$  and $\sigma_1, \sigma_2$ refer to the means and standard deviations of the two series. In the recent studies \cite{Goh2008,Zhao2011,Zhao2013,Wang2014,Zhao2016}, the memory strength coefficient $M$ was usually treated as an independent measure of the heterogeneity, with natural bounds $-1 \leq M \leq 1$.

Recently, the validity of $M$ was challenged by the sensitivity of autocorrelation function to the fat tails in $p(t)$ \cite{Karsai2012}. Indeed, if $p(t)$ follows a power-law distribution as $p(t)\sim t^{-\alpha}$ with $\alpha<2$, the autocorrelation function $A(\tau)$, defined as the Pearson correlation coefficient between $ \{t_{1}, t_{2}, \cdots, t_{n-\tau} \}$ and $ \{t_{1+\tau}, t_{2+\tau}, \cdots, t_{n}\} $ (i.e., $M=A(1)$), also decays in a power-law form as $A(\tau)\sim \tau^{-\beta}$ with $\beta=2-\alpha$. \cite{Rybski2012,Vajna2013}. Very similar to the memory strength coefficient $M$, the Pearson correlation coefficient is also applied in quantifying the degree-degree correlation in complex networks, named assortativity coefficient \cite{Newman2002,Newman2003}. Analogously, in networks with heterogeneous degree distribution, this commonly used coefficient is questioned for its nontrivial bounds in real networks \cite{Zhou2007} and theoretical network models \cite{Dorogovtsev2010,Menche2010}, as well as the dependence on network size (i.e., the assortativity coefficient decreases with the network size) \cite{Dorogovtsev2010,Raschke2010}. Accordingly, some alternative ranking-based coefficients are proposed, such as the Kendall-Gibbons' Tau \cite{Raschke2010} and Spearman's Rho \cite{Litvak2013,Hofstad2014,Zhang2016}. At the same time, the memory strength effects in human temporal activities have considerable impacts on epidemic processes \cite{Karsai2011,Min2011} and the degree-degree correlation largely affects dynamics upon the networks, including epidemic spreading \cite{Pastor2014}, evolutionary game \cite{Szabo2007}, synchronization \cite{Arenas2008}, and so on. In these works, the memory strength and correlation are quantified by the above debatable coefficients. Therefore, the understanding of fundamental properties of such coefficients in heterogeneous systems is very valuable.

In this paper, given the elements $t_1,t_2,\cdots,t_n$, we calculate the maximum and minimum values of $M$ under any permutation, suggesting the dependence between memory strength coefficient and series heterogeneity. In particular, given a power-law distribution $p(t)\sim t^{-\alpha}$, we show unreported nontrivial bounds of $M$ in the thermodynamic limit (i.e., $ n \rightarrow \infty $) --- when $1 < \alpha \leq 3$, as $\alpha$ grows bigger, the upper bound increases from 0 to +1 while the lower bound remains 0; when $\alpha > 3$, the upper bound remains +1 while the lower bound descends below 0 but strictly above -1. The theoretical bounds agree very well with numerical simulations. In addition, according to the empirical analysis on \emph{MovieLens} and \emph{Twitter}, power-law distributed inter-event time series produced by human activities are found to conform to such bounds. Our findings add novel insights in characterizing not only heterogeneous time series, but also heterogeneous networks and other complex systems with heavy-tailed distributions.

The rest of this paper is organized as follows. In Section II, we will derive the theoretical bounds of $M$ given $p(t) \sim t^{-\alpha}$. Such bounds will get validated based on extensive numerical tests in Section III and empirical results in Section IV. We will draw the main conclusion, discuss the relevance and implication of our findings in Section V.

\section{Theoretical Bounds}

Given a finite set of real numbers $\{t_1,t_2,\cdots,t_n\}$ independently sampled from a certain distribution, without loss of generality, we order them as $t_1 \leq t_2 \leq \cdots \leq t_n$, which are referred to as \textit{order statistics}. Then, by applying a permutation $ \theta $ (a one-to-one mapping from the set $ \{1, 2, \cdots, n\} $ to itself) to $ \{t_i\} $, we have a new sequence $ \{ t_{\theta_{1}}, t_{\theta_{2}}, \cdots, t_{\theta_{n}}\} $ with a different interdependence structure among elements. By changing $ \theta $, we can expect series with different values of $M$. For example, if series are permuted such that big elements tend to be followed by big ones and small followed by small ones, $ M $ would be positive; on the contrary, if big elements followed by small ones and small elements followed by big ones, $M$ would be negative. Interestingly, there exists explicit $ \theta_{\max} $ and $ \theta_{\min} $ (though not unique) that respectively maximizes and minimizes $ M $ among all possible permutations. And we can use these two extremes to derive the bounds for memory strength in the sense of all permuted independent samples.

To see this, we need to find out how a permutation affects $ M $. As the values of the two aforementioned series are different in only one element ($ t_{1} $ in head and $ t_{n} $ in tail), we assume $ m_{1} = m_{2} = m$ and
$ \sigma_{1} = \sigma_{2} = \sigma $ when $ n $ is large, where $ m $ and $ \sigma $ are the mean and standard deviation of the whole series. The memory strength of $ \{t_{\theta_i}\} $ can hence be rewritten as
\begin{equation}
	M(t_\theta) = \frac{1}{\sigma^2} \left( \frac{1}{n-1} \sum_{i=1}^{n-1} t_{\theta_i}t_{\theta_{i+1}} - m^2 \right), \label{eqs:MSimple}
\end{equation}
where the reordering of the series only affects the summed products of adjacent terms
\begin{equation}
 S_{\theta} = \sum_{i=1}^{n-1} t_{\theta_i} t_{\theta_{i+1}}, \label{summed products}
\end{equation}
while $ m $ and $ \sigma $ are invariant to permutations. The desired extreme permutations for $ M $ are just those maximize/minimize $ S_{\theta} $, denoted by $ \theta_{\max} $ and $ \theta_{\min} $ respectively. It has been shown~\cite{Hallin1992} that, for any $ n $,  there are explicit solutions to $ \theta_{\max} $ and $ \theta_{\min} $ for any real numbers $ t_{1} \leq t_{2} \leq \cdots \leq t_{n} $. Notice that, whereas Hallin \emph{et al.} \cite{Hallin1992} use an objective function that sums the products in a cycle (and therefore they call the problem \textit{optimal Hamiltonian cycles}), i.e. $S^{\prime}_{\theta} = \sum_{i=1}^{n-1} t_{\theta_{i}} t_{\theta_{i+1}} + t_{\theta_{1}} t_{\theta_{n}}$, the results can be reduced to our case by introducing an additional element $ t_0 = 0 $ to the series, which makes zero contribution to the sum.

$ \theta_{\max} $ achieves the maximum memory strength by first arranging the odd elements of order statistics in the increasing order, followed by even elements in the decreasing order, which is
\begin{equation}
\begin{split}
&t_{1}, t_{3}, \cdots, t_{2l-1}, t_{2l}, t_{2l-2}, \cdots, t_{4}, t_{2} \quad (n=2l), \\
&t_{1}, t_{3}, \cdots, t_{2l-1}, t_{2l+1}, t_{2l}, \cdots, t_{4}, t_{2} \quad (n=2l+1). \\
\end{split} \label{eqs:maxArrange}
\end{equation}
For simplicity, we only address the case when $ n=2l $, the sum with order statistics is expressed as
\begin{equation}
	S_{\theta_{\max}} = \sum_{i=1}^{2l-2} t_{i}t_{i+2} + t_{2l}t_{2l-1} \quad (n=2l), \label{eqs:Smax}
\end{equation}
while the case of $ n=2l+1 $ can be handled analogously.

On the contrary, $ \theta_{\min} $ arranges the order statistics by alternating small and big terms, namely
\begin{equation}
\begin{split}
&t_{2l}, t_{1}, t_{2l-2}, \cdots, t_{2l-3}, t_{2}, t_{2l-1} \quad (n=2l), \\
&t_{2l}, t_{2}, \cdots, t_{l}, \cdots, t_{1}, t_{2l+1} \quad (n=2l+1),
\end{split} \label{eqs:minArrange}
\end{equation}
where when $n$ is even, even and odd terms are interlaced; when $n$ is odd, half of the sequence is made up of even terms while the other half odd.
Similarly, for even $ n $, we have
\begin{equation}
	S_{\theta_{\min}} = \sum_{i=1}^{l-1} \big( t_{i}t_{2l+1-i} + t_{i}t_{2l-1-i} \big) + t_{l}t_{l+1} \quad (n=2l),
\end{equation}
and the case of $n=2l+1$ is analogous.

We then define the upper bound and lower bound for memory strength as the expectation of $ M $ under $ \theta_{\max} $  and $ \theta_{\min} $ in the limit of infinitely long series. The bounds, defined as
\begin{equation}
M_{\max} = \E \lim_{n \rightarrow \infty} M(t_{\theta_{\max}})
\end{equation}
and
\begin{equation}
M_{\min} = \E \lim_{n \rightarrow \infty} M(t_{\theta_{\min}}),
\end{equation}
measure the memory strength constraints imposed by a marginal distribution, where E denotes the operator to obtain the expected value. As we will see, these bounds can be derived in a closed form or effectively approximated for several distributions.

It can be shown that $ M_{\max}=1 $ and $ M_{\min}=-1 $ hold for uniform and Gaussian distributions, corresponding to the natural range of $ M $ (see details in \emph{Appendix A} and \emph{Appendix B}). However, a much narrower range is found for power-law distributions, where the bounds rely on the exponent $ \alpha $. While the rest of the paper focuses on power-law, we also notice a few other distributions with non-trivial memory strength constraints, which are discussed in \emph{Appendix C}.

Supposing that the series $\{t_1,t_2,\cdots,t_n\}$ are independently sampled from a power-law distribution with density
\begin{equation}
p(t)= (\alpha-1) t^{-\alpha} \quad (t \geq 1).
\end{equation}
To derive how the bounds rely on $ \alpha $, we first consider the case where $ \alpha > 3 $, which is necessary for the population variance $ \sigma({\alpha})^{2} $ (appearing in the denominator of $ M $) to converge.
When $ \alpha>3 $, by the strong law of large numbers and continuous mapping theorem, we have
\begin{equation*}
M(t_\theta) \rightarrow \frac{1}{\sigma(\alpha)^2} \left(\lim_{n \rightarrow \infty} \frac{1}{n-1} \sum_{i=1}^{n-1} t_{\theta_i}t_{\theta_{i+1}} -m(\alpha)^2 \right)
\end{equation*}
almost surely as $n \rightarrow \infty$, where the sample moments are replaced by the corresponding population moments
\begin{equation}
m(\alpha) = \int_{1}^{+\infty} tp(t) dt = \frac{\alpha-1}{\alpha-2}, \label{eqs:m-alpha}
\end{equation}
and
\begin{equation}
\sigma(\alpha)^2 = \int_{1}^{+\infty} (t-m(\alpha))^{2}p(t) dt= \frac{\alpha-1}{\alpha-3} - \left(\frac{\alpha-1}{\alpha-2}\right)^2. \label{eqs:sigma-alpha}
\end{equation}
Here we assume $ t_{\min}=1 $ because $ M $ would remain the same if every $ t_{i} $ is divided by the same constant. Since $|M| \leq 1$, by Lebesgue's dominated convergence theorem, we can switch the order between limit and expectation and have
\begin{equation}
\E \lim_{n \rightarrow \infty} M(t_{\theta}) = \frac{1}{\sigma(\alpha)^2} \left(\lim_{n \rightarrow \infty} \frac{1}{n-1} \E S_{\theta} -m(\alpha)^2 \right). \label{eqs:Mprob}
\end{equation}
Therefore, in the $\alpha>3$ case, $M_{\max}$ and $M_{\min}$ can be respectively determined by $\frac{1}{n-1} E S_{\theta_{\max}}$ and $\frac{1}{n-1} E S_{\theta_{\min}}$ in the limit of large $n$.

According to Eq.~(\ref{eqs:Smax}), for the case of $n=2l$, we have
\begin{equation}
\frac{1}{n-1}\E S_{\theta_{\max}} = \frac{1}{2l-1}\sum_{i=1}^{2l-2} \E t_{i}t_{i+2} + \frac{\E t_{2l} t_{2l-1}}{2l-1}, \label{eqs:s-maxmem}
\end{equation}
whereas the case of $ n=2l+1 $ can be worked out in a similar fashion to arrive at the same result.

The expected value of each term can be obtained by using the joint distribution of order statistics.
The probability density function for the joint distribution of two order statistics $ t_{j}, t_{k}\  (j<k)$ is given by \cite{David2003}
\begin{equation}
\begin{split}
p(t_j=x,t_k=y) &= n! \frac{[P(x)]^{j-1}}{(j-1)!} \frac{[P(y) - P(x)]^{k-1-j}}{(k-1-j)!} \\
& \times \frac{[1-P(y)]^{n-k}}{(n-k)!} p(x) p(y) \quad  (x \leq y), \label{eqs:joint-order-stat}
\end{split}
\end{equation}
where $ P(x) $ is the corresponding cumulative distribution function for power-law, i.e.
\begin{equation*}
	P(x) = \int_{1}^{x}p(t)dt = 1 - x^{1-\alpha}.
\end{equation*}
Therefore, we have
\begin{equation*}
\begin{split}
	\E t_{i} t_{i+2} &= \iint_{1 \leq x \leq y < \infty} xy p(t_i=x,t_{i+2}=y) dxdy \\
	&= \frac{\Gamma(2l+1)}{\Gamma(2l+1-2c)} \frac{\Gamma(2l-i+1-2c)}{\Gamma(2l-i-1)} \\
	& \times \frac{1}{(2l-i-c)(2l-i-\alpha c)}
\end{split}
\end{equation*}
for $1 \leq i \leq 2l-2$ and the boundary term
\begin{equation*}
\begin{split}
\E t_{2l}t_{2l-1} &= \iint_{1 \leq x \leq y < \infty} x y f_{t_{(2l-1)}, t_{(2l)}}(x,y) dx dy \\
&= \frac{\alpha-1}{\alpha-2} \frac{\Gamma(2-2c)\Gamma(2l+1)}{\Gamma(2l+1-2c)},
\end{split}
\end{equation*}
where a shorthand $ c = \frac{1}{\alpha-1}$ ($ 0<c<\frac{1}{2} $) is adopted.

In the limit of $ n \rightarrow \infty $, the first term in (\ref{eqs:s-maxmem}) would be
\begin{equation}
\begin{split}
& \quad \lim_{l \rightarrow \infty} \frac{1}{2l-1} \sum_{i=1}^{2l-2} \E t_{i} t_{i+2} \\
&= \lim_{l \rightarrow \infty} (2l+1)^{-(1-2c)} \sum_{k=2}^{2l-1} \frac{1}{(k-c)(k-c-1)} \frac{\Gamma(k+1-2c)}{\Gamma(k-1)} \\
&= \lim_{l \rightarrow \infty} \sum_{k=2}^{2l-1} (\frac{k}{2l+1})^{-2c} \frac{1}{2l+1}= \int_{0}^{1} t^{-2c} dt = \frac{\alpha-1}{\alpha-3}, \label{eqs:s1-maxmem}
\end{split}
\end{equation}
where we have applied the property $\lim_{x \rightarrow \infty} \frac{\Gamma(x) x^\gamma}{\Gamma(x+\gamma) } = 1$ for real $\gamma$, and rewrote the summation with $k = 2l - i$.

Meanwhile, the boundary term in the right hand side of Eq. (\ref{eqs:s-maxmem}) vanishes in the limit of large $ n $, i.e.
\begin{equation}
	\lim_{l \rightarrow \infty} \frac{1}{2l-1} \E t_{2l} t_{2l-1} = 0. \label{eqs:s2-maxmem}
\end{equation}
Substituting Eqs. (12), (13), (\ref{eqs:s1-maxmem}) and (\ref{eqs:s2-maxmem}) into (\ref{eqs:Mprob}), we arrive at
\begin{equation}
	M_{\max} =  \E \lim_{n \rightarrow \infty} M_{\theta_{\max}} = 1 \quad (\alpha>3). \label{eqs:max-mem-theory}
\end{equation}

\begin{figure}[!htp]
\includegraphics[width=0.9\columnwidth]{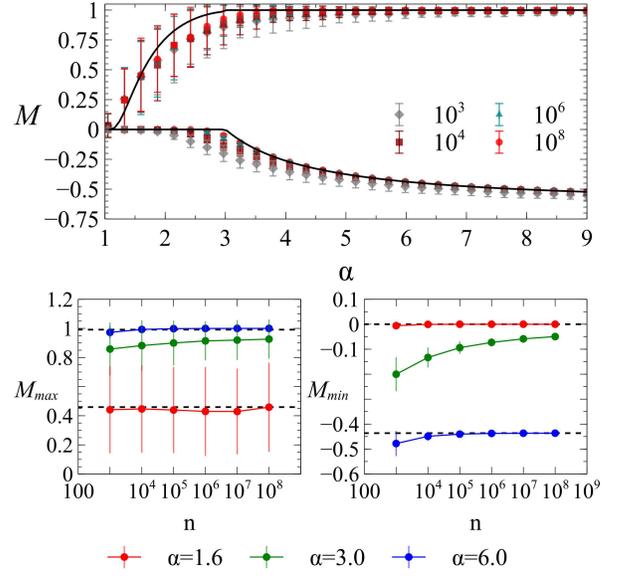}
\caption{(Color online) Top: (a) Theoretical bounds (black solid) compared to simulated sample mean (dots) of $M(t_{\theta_{\max}})$ and $M(t_{\theta_{\min}})$ under different sequence lengths $n=10^{3}, 10^{4}, 10^{6}$ and $10^{8}$. Each dot is averaged over 1,000 independent draws of the sequence $\{t_{i}\}$ and the error bar denotes one standard deviation. Below: Finite-size scaling analysis of the (b) maximum and (c) minimum values of $M$ as $n$ increases, under $\alpha = 1.6, 3.0$ and $6.0$. Mean values approach their corresponding theoretical bounds (dashed lines) as $n$ increases.}
\label{fig:bounds}
\end{figure}

Similarly, to obtain the lower bound when $\alpha>3$, we rewrite the summed products as
\begin{equation*}
\begin{split}
\frac{1}{n-1} \E S_{\theta_{\min}} = &\frac{1}{2l-1} \sum_{i=1}^{l-1} \left(\E t_{i} t_{2l+1-i} + \E t_{i} t_{2l-1-i} \right) + \\
& \frac{1}{2l-1} \E t_{l}t_{l+1}.
\end{split}
\end{equation*}
Again, we assume $ n=2l $ for convenience, while the case of $ n=2l+1 $ gives the same result. Taking the large $ n $ limit, we have
\begin{equation*}
\begin{split}
&\quad \lim_{l \rightarrow \infty} \frac{1}{2l-1} \sum_{i=1}^{l-1} \E t_{i} t_{2l+1-i} \\
&= \lim_{l \rightarrow \infty} (2l+1)^{-(1-2c)} \sum_{i=1}^{l-1} (i+2-c)^{-c} (2l-i+1-2c)^{-c} \\
&= \lim_{l \rightarrow \infty} \frac{1}{2l-1} \sum_{i=1}^{l-1} \left( \frac{i}{2l+1} + \frac{2-c}{2l+1} \right)^{-c} \left( 1- \frac{i+2c}{2l+1} \right)^{-c} \\
&= \lim_{l \rightarrow \infty} \frac{1}{2l-1} \sum_{i=1}^{l-1} \left( \frac{i}{2l+1} \right)^{-c} \left( 1-\frac{i}{2l+1}\right)^{-c}\\
&= \int_{0}^{\frac{1}{2}} u^{-c} (1-u)^{-c} du = B \left(\frac{1}{2}; \frac{\alpha-2}{\alpha-1}, \frac{\alpha-2}{\alpha-1} \right),\\
\end{split}
\end{equation*}
where $u=i/(2l+1)$ and $B(\cdot)$ is the \textit{incomplete beta function}
\begin{equation*}
	B(x; a,b) = \int_{0}^{x} t^{a-1} (1-t)^{b-1} dt.
\end{equation*}
The other term has the same limit, namely
\begin{equation*}
\lim_{l \rightarrow \infty} \frac{1}{2l-1} \sum_{i=1}^{l-1} \E t_{i} t_{2l-1-i} = B \left( \frac{1}{2}; \frac{\alpha-2}{\alpha-1}, \frac{\alpha-2}{\alpha-1} \right),
\end{equation*}
and again the remaining term vanishes as
\begin{equation*}
	\lim_{l \rightarrow \infty} \frac{1}{2l-1} \E t_lt_{l+1} = 0.
\end{equation*}
Therefore, the minimum memory strength for $\alpha>3$ is
\begin{equation}
	M_{\min} = \frac{1}{\sigma(\alpha)^2} \left[ 2B\left(\frac{1}{2}; \frac{1}{m(\alpha)}, \frac{1}{m(\alpha)}\right)  - m(\alpha)^2\right], \label{eqs:min-mem-theory}
\end{equation}
where the population moments are given by Eqs. (\ref{eqs:m-alpha}) and (\ref{eqs:sigma-alpha}). Noticeably, $M_{\min}$ is a decreasing function of $\alpha$ (see Figure \ref{fig:bounds}) that approaches -0.65 as $\alpha$ gets large.

Having obtained the upper bound (\ref{eqs:max-mem-theory}) and the lower bound (\ref{eqs:min-mem-theory}) for $ \alpha>3 $, we now turn to the case for $ 1 < \alpha \leq 3 $ ($ \alpha > 1$ is necessary for power-law to be normalized). In this case, the corresponding population moment for $ \sigma^{2} $ would diverge ($ m(\alpha) $ also diverges when $ \alpha <2 $), rendering the moment-substitution technique infeasible. Meanwhile, it seems formidable to deal with the probability distributions of $ M(t_{\theta_{\max}}) $ and $ M(t_{\theta_{\min}}) $ directly. Therefore, we present an approximation method that recovers the asymptotic behavior of statistics when $ n \rightarrow \infty $ by substituting random variables with deterministic surrogates.

To do so, we pick the points $ \{\hat{t}_1, \hat{t}_2, \cdots, \hat{t}_n\} $ that cut the area under the probability density function $ p(t) $ into slices of equal area $ 1/n $, with $ \hat{t}_1 = x_{\min} = 1 $ and the area from $ t_{n} $ extending to infinity also being $ 1/n $. Then we approximate the random samples $ \{ t_{(1)}, t_{(2)}, \cdots, t_{(n)} \} $  with these deterministic points $ \{ \hat{t}_1, \hat{t}_2, \cdots, \hat{t}_n \} $. It should be noted that such approximation imposes a cut-off on the maximum value of $ \{t_i\} $ and the probability of drawing a sample exceeding the cut-off is $ 1/n $ , which diminishes to zero as $ n \rightarrow \infty $.
As $ \int_{1}^{\hat{t}_i} p(t)dt = (i-1)/n $, we have
\begin{equation}
	\hat{t}_i = (1-\frac{i-1}{n})^{-c}  \quad (i=1,2,\cdots,n),
\end{equation}
where $ c = 1/(\alpha-1) > \frac{1}{2} $ in this case.

Rewriting $ M $ in terms of samples as
\begin{equation}
	M = \frac{s-m^{2}}{\overline{t^{2}} -m^{2}},
\end{equation}
where $ s = \frac{1}{n-1} S = \frac{1}{n-1} \sum_{i=1}^{n-1} t_{i} t_{i+1}$, $ \overline{t^2} = \frac{1}{n} \sum_{i=1}^n t_i^2 $ and $ m^2 = (\frac{1}{n}\sum_{i=1}^n t_i)^2 $ are the statistics in concern. By substituting $t_i$ with $ \hat{t}_i $, we seek approximations for these statistics (denoted by $\hat{s}$, $\overline{\hat{t}^{2}}$ and $\hat{m}^{2}$) in the form of $ n^{\mu(\alpha)}g(n,\alpha) $, where when $ n \rightarrow \infty $, $ g(n,\alpha) $ converges to a non-zero function of $ \alpha $ while the divergence is characterized by the polynomial term $ n^{\mu(\alpha)} $. This family of functions are denoted by $O(n^{\mu(\alpha)})$ generically. With some algebra, we have
\begin{equation*}
\hat{m}^{2} = \begin{cases} O(1) \quad &(2 < \alpha < 3) \\
n^{2c-2} (\sum_{k=1}^{n} k^{-c})^{2}=O(n^{2c-2}) \quad &(1 < \alpha \leq 2)
\end{cases}
\end{equation*}
and
\begin{equation*}
\overline{\hat{t}^{2}} = n^{2c-1}\sum_{k=1}^{n} k^{-2c} = O(n^{2c-1}),
\end{equation*}
which two hold for both $\theta_{\max}$ and $\theta_{\min}$.

Then, for the upper bound, we have
\begin{equation}
\hat{s}_{\theta_{\max}} = \frac{n^{2c}}{n-1} \big[ \sum_{k=2}^{n-1} (k^2-1)^{-c} + 2^{-c} \big],
\end{equation}
which diverges with the order of $ O(n^{2c-1}) $. By comparing it with the order of $\hat{m}^{2}$ and $\overline{\hat{t}^{2}}$, in the limit of large $ n $, we know $ M_{\theta_{\max}} $ can be approximated by neglecting $ \hat{m}^2 $. Therefore, for $1 < \alpha \leq 3$ we have
\begin{equation}
M_{\max} \approx \lim_{n \rightarrow \infty} \hat{M}_{\theta_{\max}} = \lim_{n \rightarrow \infty} \frac{\sum_{k=2}^{n-1} (k^2-1)^{-c} + 2^{-c}}{\sum_{k=1}^{n} k^{-2c}},
\label{eqs:mmax-approximate-theory}
\end{equation}
where both the numerator and the denominator are convergent and can thus be approximately computed by taking a large $ n $.

Meanwhile, supposing $n=2l$ for convenience, we have
\begin{equation*}
\begin{split}
&\hat{s}_{\theta_{\min}} = \frac{n^{2c}}{n-1} \Big[ \frac{1}{2}\sum_{i=1}^{n}[i(n+1-i)]^{-c} \\
&+ \frac{1}{2}\sum_{i=1}^{n}[(i+2)(n+1-i)]^{-c} -(n+2)^{-c}/2 - (2n+2)^{-c}/2 \Big] .
\end{split}
\end{equation*}
By observing when $c>1/2$
\begin{equation*}
\sum_{i=1}^{n} [i(n+1-i)]^{-c} \leq \sum_{i=1}^{n} (n+1)^{-2c} = n (n+1)^{-2c} \rightarrow 0
\end{equation*}
and similarly for the other sum, we know $\hat{s}_{\theta_{\min}} = o(n^{2c-1})$, which diverges more slowly than $\overline{\hat{t}^{2}}$. Here $ f(n) = o(g(n))$ means that $ f(n)/g(n) \rightarrow 0$ in the limit of large $n$. Because the term with the biggest order only appears in the denominator, we have
\begin{equation}
	M_{\min} \approx \lim_{n \rightarrow \infty} \hat{M}_{\theta_{\min}} = 0
\end{equation}
for $1 < \alpha \leq 3$. This non-negative constraint is particularly interesting because many power-law series are empirically found to have $\alpha$ in this region~\cite{Oliveira2005,Eagle2005,Vazquez2006,Dezso2006,Lambiotte2007,Zhou2008,Wang2008,Li2008,Goncalves2008,Baek2008,Hong2009,Radicchi2009,Wu2010,Wang2011,Takaguchi2011,Zhou2012,Zhao2012b,Kondor2014,Picoli2014,Hou2014,Zha2015}.

\begin{figure}[!htp]
\includegraphics[width=0.99\columnwidth]{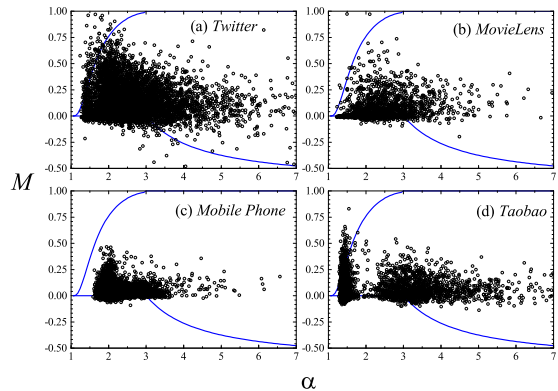}
\caption{(Color online) Memory strength and fitted $\alpha$ for power-law distributed inter-event time series produced by (a)posting to \emph{Twitter}, (b) rating on \emph{MovieLens}, (c) phoning by Orange's \emph{Mobile Phone}, and (d) browsing on \emph{Taobao}. Each circle corresponds to a user. Theoretical bounds are drawn by blue solid line. }
\label{fig:empirical}
\end{figure}

\section{Simulations}

In Figure \ref{fig:bounds}, we compare theoretical bounds with simulated $M(t_{\theta_{\max}})$ and $M(t_{\theta_{\min}})$ under difference lengths $n$ of the sequence $\{t_1,t_2,\cdots,t_n\}$. First, we observe that with a larger $n$, the theoretical bounds, defined as the expectation under infinite $n$, match the sample mean of $M(t_{\theta_{\max}})$ and $M(t_{\theta_{\min}})$ more closely. Second, in terms of accuracy, the effect of approximation is noticeable but satisfactory in the upper bound when $1 < \alpha \leq 3$; meanwhile, the lower bound and the upper bound for $\alpha>3$ are both accurately predicted. Third, because $M(t_{\theta_{\max}})$ and $M(t_{\theta_{\min}})$ are random variables for each $\alpha$, we quantify their variance around the mean in terms of standard deviation in the figure. Although our theoretical bounds do not predict their variance, we speculate that as $n$ tends to infinity, $M(t_{\theta_{\min}})$ converges almost surely to the deterministic theoretical lower bounds; similarly, $M(t_{\theta_{\max}})$ converges to constant 1 for $\alpha >3$ because $M \leq 1$. However, when $1<\alpha \leq 3$, $M(t_{\theta_{\max}})$ converges to a non-degenerate random variable whose distribution depends on $\alpha$, as evidenced by the non-vanishing error bars in that region.

\section{Empirical Results}
In this section, we examine the empirical distribution of $(M, \alpha)$ of power-law distributed sequences to see if real physical processes conform to the memory strength constraints predicted by our theory. To this end, we use inter-event time series collected from online human activities \cite{Zhao2012b}. Inter-event time series refer to the series made up of time intervals between every two consecutive events and have been widely found to follow power-law distributions~\cite{Barabasi2005}.

Figure 2 reports findings from \emph{Twitter}, \emph{MovieLens}, \emph{Mobile Phone} and \emph{Taobao} datasets (each circle represents an individual user) compared to theoretical bounds (blue solid). Similar graphs have been proposed as a ``phase diagram'' in which different systems are grouped into different regions~\cite{Goh2008}. The \emph{Twitter} dataset (a year-long subset of tweets crawled by Choudhury \emph{et al.}~\cite{DeChoudhury2010}, starting from Nov 2008) collects the time stamps from 9,832,781 tweets posted by 117,436 users. And, the series correspond to the time intervals between two consecutive tweeting in this dataset. \emph{MovieLens} is a website where users rate movies and get recommendations based on their ratings. The \emph{MovieLens} 10M dataset collects $10^7$ time stamps from 71,567 users when they rate a movie online (the collection of \emph{MovieLens} data started years ago~\cite{Resnick1994} and can be freely downloaded from http://grouplens.org/). In this dataset, the series correspond to the time intervals between two consecutive rating. The \emph{Mobile Phone} dataset comes from the Orange ``Data for Development'' (D4D) challenge~\cite{Vincent2012}, which is an open data challenge on anonymous call patterns of Orange's mobile phone users in Ivory Coast. Four mobile phone datasets are accessible through this challege, and the data we used in this paper is specifically the file ``SUBPREF\underline{\hspace{0.5em}}POS\underline{\hspace{0.5em}}SAMPLE\underline{\hspace{0.5em}}A.TSV'' in the archive \emph{SET3}. In this dataset, the calling records of 500,000 randomly selected individuals are provided, and the series correspond to the time intervals between two consecutive phone calls. \emph{Taobao}, a Chinese web site, is one of the world's largest electronic marketplaces and our data is composed of all browsing behaviors of 34,330 users, each has visited more than 100 items in the time span between September 1 and October 28, 2011~\cite{Zhao2012b}. We study the time series of browsing behavior of individuals in this paper.

For comparison, we rule out the series in the datasets that are either too short ($n<180$ for \emph{Twitter}, $n<200$ for \emph{MovieLens} and \emph{Taobao}, and $n<300$ for \emph{Mobile Phone}) or are unlikely to follow a power-law ($p\text{-value}>0.1$). The $p$-value is computed from a goodness-of-fit test based on the Kolmogorov-Smirnov statistic as suggested by \cite{Clauset2009}. As a result, we have 5,517 series from \emph{Twitter}, 2,261 series from \emph{MovieLens}, 24,519 series from \emph{Mobile Phone}, and 5,107 series from \emph{Taobao}, of which $\alpha$ is estimated with maximum likelihood \cite{Clauset2009}. As can be seen from the plot, except for a few outliers, most of the sequences fall into the predicted region. It is worth recapitulating that the predicted upper bound for $1<\alpha<3$ is an approximate to the mean, and the non-zero variance of $M(t_{\theta_{\max}})$ (see Figure 1) allows some points to exceed the predicted upper bound. Notice that, the theoretical bounds are derived under two strong conditions: the target sequence is infinitely long, with its elements following a prefect power-law distribution. In contrast, real sequences are very short ($n \sim 10^2$ for real sequences, while in figure 1, we have tested the case with $n=10^8$) and far from perfect power laws. Therefore, the results presented in Fig. 2 indicates the high applicability of the theoretical bounds to real heterogenous systems.

\section{Conclusion and Discussion}
In this article, we explore non-trivial bounds on the memory strength of power-law distributed series, which challenge the common treatment of memory strength as a measure independent of heterogeneity. We seek the bounds inside an ensemble formed by permuting independently and identically distributed power-law series, which covers a wide range of dependent structures while preserving the marginal distribution. Based on results in permutational extreme values, we present the bounds in either closed form or with an effective approximation.

Our bounding technique relies on a permutation ensemble of which the extremes are solvable. Although it does not subsume all possible power-law distributed series, we speculate that it is reasonably flexible that is covers most ``natural'' series. For example, series with $t_1 \sim p(x)$ and $t_1=t_2=\cdots=t_n$ is an ``unnatural'' construction that is not contained by the ensemble. It is worth further investigation on the applicability of the permutation ensemble and the possibility of constructing other tractable ensembles with larger flexibility and fewer assumptions. Nevertheless, the usefulness of permutation ensemble is evidenced by empirical data (human activities in this case) found to closely conform to such bounds.

Without the knowledge of the present non-trivial bounds of memory strength, one may straightforwardly think that the values of $M$ in the permutation ensemble are uniformly distributed in the range $[-1,1]$ with average value being equal to 0. If so, when one observes a positive value of $M$ as in many real human-activated systems~\cite{Goh2008}, he/she will feel that the memory strength effect of this time series is stronger than the average of the permutation ensemble. Analogously, when one observes a positive assortativity coefficient in a network, he/she will feel that the large-degree nodes tend to connect with other large-degree nodes, compared with the randomized networks with the same degree sequence. However, as indicated by the present results, the above intuition is incorrect. For example, for a power-law time series with exponent $\alpha$ in the range $(1,3]$, the memory strength $M$ is non-negative so that a small positive value of $M$ is possibly still \emph{smaller} than the average of the permutation ensemble. That is to say, given the values of elements in the time series, a high-value element is indeed more likely to be followed by some lower-value elements than the randomized counterpart even for some $M>0$. Analogously, in a scale-free network~\cite{Barabasi1999} with power-law exponent $\gamma$, an extension of the present method yields a nontrivial lower bound of assortativity coefficient $r_{\min}$ in the limit of network size as (details will be published elsewhere):
\begin{equation}
r_{\min} =
\begin{cases}
0 \quad &(2 < \gamma \leq 4) \\
\frac{2B\left( \frac{1}{2}; \frac{\gamma-3}{\gamma-2}, \frac{\gamma-3}{\gamma-2}\right)-\left( \frac{\gamma-2}{\gamma-3} \right)^2}{\frac{\gamma-2}{\gamma-4} - \left( \frac{\gamma-2}{\gamma-3}\right)^2} \quad &(\gamma > 4).
\end{cases}
\end{equation}
This bound improves the previously known results (see, for example, Eqs. (18-23) in \cite{Dorogovtsev2010} and Eq. (64) in \cite{Menche2010}), in particular for $\gamma>4$. Obviously, when $\gamma \in (2,4]$, the lower bound of assortativity coefficient is zero, hence in a real network with assortativity coefficient larger than zero, a large-degree node may be more likely to connect with some smaller-degree nodes than the average in the corresponding null networks with the same degree sequence~\cite{Maslov2002}. Therefore, bringing to light the existence of non-trivial bounds of $M$ (and possibly other statistical measures to be revealed by future studies) in heterogeneous systems, this work could clear up some misunderstanding from the ostensible values of autocorrelation, assortativity coefficient, etc. Since many power-law series have $\alpha$ in the range $(1,3]$ and many scale-free networks have $\gamma$ in the range $(2,4]$, our findings are relevant and significant to the academic society with interests in real complex systems.

Lastly, this work gives rise to a natural question, namely how to do proper statistics on heterogeneous systems, whose characteristic distributions may be mathematically ill-posed and/or have divergent finite moments, challenging the validity and usefulness of many commonly-used statistical measures. This question has been attempted by a few earlier works, among which, notably, Karsai \emph{et al.} \cite{Karsai2012} reported the problem with autocorrelation function and the Hurst exponent for characterizing heterogeneous systems, as they ``can assign false positive correlations'' even when the temporal dependence is absent or destroyed. Yet, a proper alternative measure, one with desirable theoretical properties, still remains elusive. Given a certain statistical measure computed from a system, it would be important to know the relative position of this value compared to the distribution of the same measure computed from a ``reference ensemble'' of systems, such as the permutation ensemble in this case. However, such a task is usually difficult due to the computational complexity associated with the ensemble analysis. Therefore, unfolding the mathematical structure of reference ensembles could be very helpful, which is left as our future aim.

\begin{acknowledgments}
    The authors acknowledge valuable discussion with Xiaoyong Yan, Chenmin Sun and Yifan Wu. This work is partially supported by National Natural Science Foundation of China under Grants Nos.~61433014 and 11222543. T.Z. acknowledges the Program for New Century Excellent Talents in University under Grant No. NCET-11-0070, and Special Project of Sichuan Youth Science and Technology Innovation Research Team under Grant No. 2013TD0006.
\end{acknowledgments}

\appendix

\section{Bounds for Uniform Distribution}

Since $M(\cdot)$ is invariant to translation and scaling, without loss of generality, we assume that $t_1,t_2,\cdots,t_n$ are sampled from a uniform distribution in the range $(0,1)$, namely $\{t_{i}\} \sim \text{Unif}(0,1)$. We use the same technique as in the paper to show that the theoretical bounds for uniform distributed sequence are the ordinary $\pm 1$ bounds.

\begin{figure}[!htp]
\includegraphics[width=0.85\columnwidth]{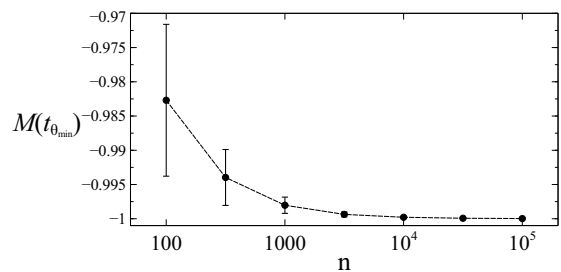}
\caption{$M(t_{\theta_{\min}})$ simulated for different $n$, where $\{t_i\}$ are sampled from a Gaussian distribution. The dot represents sample mean from 1,000 independent draws and error bar represents one standard deviation.}
\label{fig:lower-gaussian}
\end{figure}

\begin{figure*}[!htp]
\includegraphics[width=1.0\textwidth]{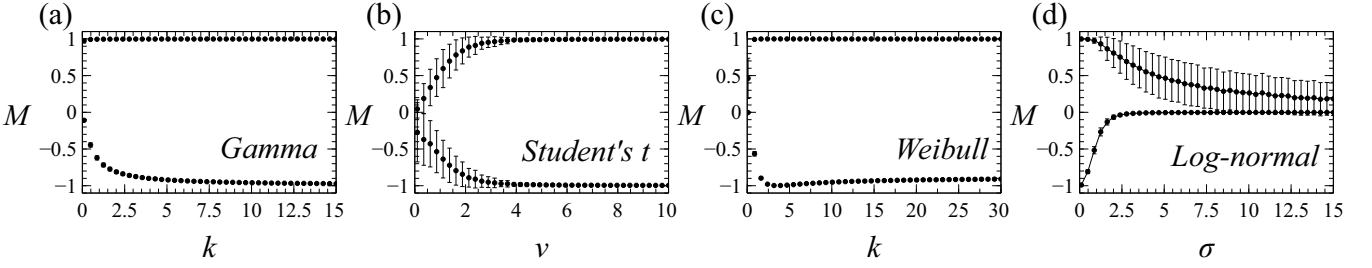}
\caption{Simulated $M(t_{\theta_{\max}})$ and $ M(t_{\theta_{\min}}) $ for various distributions also with non-trivial memory strength constraints: (a) Gamma distribution $p(x) \propto (x/\lambda)^{k-1}e^{-x/\lambda}$ with shape parameter $k$, (b) Student's t-distribution $p(x) \propto (1+x^2/\nu)^{-(\nu+1)/2}$ with degree of freedom $\nu$, (c) Weibull distribution $p(x) \propto (x/\lambda)^{k-1}e^{-(x/\lambda)^k}$ with shape parameter $k$, and (d) Log-normal distribution $p(x) \propto x^{-1} e^{-\frac{(\ln x-\mu)^2}{2\sigma^2}}$ with scale parameter $\sigma$. Each dot is averaged over 1,000 independent draws with series length $n=1,000$. Error bars represent one standard deviation.}
\label{fig:other-distributions}
\end{figure*}

Obviously $m=1/2$ and $\sigma^{2}=1/12$, and the joint distribution for two order statistics is given by
\begin{equation*}
\begin{split}
p(t_{j} = x, t_{k} = y) = \frac{n!}{(j - 1) ! (k - j - 1) ! (n - k) !} \\
\times x^{j - 1} (y - x)^{k - j - 1} (1 - y)^{n - k},
\end{split}
\end{equation*}
for $ 1 \leq j < k \leq n $ and $ x \leq y$.

$ M_{\max} $ is determined by
\begin{equation*}
\lim_{n \rightarrow \infty} \frac{1}{n-1} S_{\theta_{\max}} = \lim_{l \rightarrow \infty} \frac{1}{2l-1} \sum_{i=1}^{2l-2} \E t_i t_{i+2}.
\end{equation*}
By plugging in
\begin{equation*}
\begin{split}
\E t_{i} t_{i + 2} &= \iint_{0 \leqslant x \leqslant y
\leqslant 1} xyf (t_{i} = x, t_{i + 2} = y) dx dy \\
&= \frac{i (i+ 3)}{(2 l + 1) (2 l + 2)},
\end{split}
\end{equation*}
we have $\lim_{n \rightarrow \infty} \frac{1}{n - 1} \E S_{\theta_{\max}} =
1 / 3$ and therefore
\begin{equation}
M_{\max} = \frac{1}{\sigma^2} \left( \lim_{n \rightarrow \infty} \frac{1}{n -
1} \E S_{\theta_{\max}} - m^2 \right) = 1.
\end{equation}

$M_{\min}$ is determined by
\begin{equation*}
\begin{split}
  \lim_{n \rightarrow \infty} \frac{1}{n - 1} \E S_{\theta_{\min}}
  = \lim_{l \rightarrow \infty} \frac{1}{2 l - 1} \sum_{i = 1}^{l - 1}
  \E t_{i} t_{2 l + 1 - i} \\
  + \lim_{l \rightarrow \infty}
  \frac{1}{2 l - 1} \sum_{i = 1}^{l - 1} \E t_{i} t_{2 l - 1 -
  i}.
\end{split}
\end{equation*}
Since
\begin{equation*}
\begin{split}
  \E t_{i} t_{2 l + 1 - i} & = \iint_{0 \leqslant x
  \leqslant y \leqslant 1} xyf (t_{i} = x, t_{2 l + 1 - i} - y) dx
  dy\\
  & = \frac{i (2 l - i + 2)}{(2 l + 1) (2 l + 2)},
\end{split}
\end{equation*}
we have $\lim_{l \rightarrow \infty} \frac{1}{2 l - 1} \sum_{i = 1}^{l - 1}
\E t_{i} t_{2 l + 1 - i} = \frac{1}{12}$ and similarly $\lim_{l \rightarrow \infty} \frac{1}{2 l - 1} \sum_{i = 1}^{l - 1}
\E t_{i} t_{2 l - 1 - i} = \frac{1}{12}$. Substituting these two results, we have $\lim_{n - \infty} \frac{1}{n - 1} \E S_{\theta_{\min}} =
\frac{1}{6}$. Therefore, we arrive at
\begin{equation}
M_{\min} = \frac{1}{\sigma^2} \left( \lim_{n \rightarrow \infty} \frac{1}{n -
1} \E S_{\theta_{\min}} - m^2 \right) = - 1.
\end{equation}

\section{Bounds for Gaussian Distribution}
In the case of Gaussian distributions, since the joint distribution in Eq. (\ref{eqs:joint-order-stat}) does not have an analytical form, our previous technique that evaluates the expectation of $S_{\theta_{\max}}$ and $S_{\theta_{\min}}$ becomes infeasible. Therefore, we will instead derive the upper bound for Gaussian sequences with a bounding technique, and resort to simulations to observe the lower bound.

As $M(\cdot)$ is invariant to translation and scaling, we can consider $\{t_i\} \sim N(\mu, \sigma^2)$ for any $\mu$ and $\sigma^2$. In particular, to get rid of the signs, we let $\mu$ depend on $n$ such that $\mu$ is large enough to ensure all samples to be non-negative ($\mu$ grows like $O(\sqrt{\log n})$ \cite{Coles2001}). Following Eq. (\ref{eqs:MSimple}), $ M(t_{\theta_{\max}}) $ is rewritten as
\begin{equation}
M(t_{\theta_{\max}}) = \frac{n}{n-1} \frac{S_{\theta_{\max}} - (n-1)m^2}{\sum_{i=1}^n t_i^2 - n m^2}, \label{eqs:M_Gaussian}
\end{equation}
where $m = \frac{1}{n}\sum_{i=1}^n t_i$. Assuming $n=2l$, since all samples are non-negative, by the ordering $t_1 \leq t_2 \leq \cdots \leq t_n$ we have
\begin{equation*}
\begin{split}
S_{\theta_{\max}} &= \sum_{i=1}^{2l-2} t_i t_{i+2} + t_{2l} t_{2l-1} \\
& \geq \sum_{i=1}^{2l-2} t_i^2 + t_{2l-1}^2 = \sum_{i=1}^n t_i^2 - t_n^2.
\end{split}
\end{equation*}
Now dividing both numerator and denominator by $\sum_{i=1}^n t_i^2$ in Eq. (\ref{eqs:M_Gaussian}), we have
\begin{equation}
M(t_{\theta_{\max}}) \geq \frac{n}{n-1} \frac{1 - \frac{t_n^2}{\sum_{i=1}^n t_i^2} - \frac{n-1}{n} \frac{m^2}{\frac{1}{n}\sum_{i=1}^n t_i^2}}{1- \frac{m^2}{\frac{1}{n} \sum_{i=1}^n t_i^2}},
\end{equation}
where in the right hand side, $\frac{m^2}{\frac{1}{n}\sum_{i=1}^n t_i^2} \rightarrow \frac{\mu^2}{\sigma^2 + \mu^2}$ almost surely. Also, it has been shown that the ratio $ {t_n^2}/{\sum_{i=1}^n t_i^2}$ between the largest element and the sum converges to 0 almost surely for independent sequence $\{t_i^2\}$ drawn from a common distribution with finite expectation \cite{OBrien1980, Downey2007}. Hence, the right hand side of Eq.~(\ref{eqs:M_Gaussian}) converges to 1 almost surely. Meanwhile, we know $M(t_{\theta_{\max}}) \leq 1$. Therefore, we know the bounded sequence $M(t_{\theta_{\max}}) \rightarrow 1$ almost surely and thus $M_{\max}=1$.

We conjecture that $M(t_{\theta_{\min}}) \rightarrow -1$ almost surely also holds for the lower bound, although a proof is not obvious. In support of this result, Figure \ref{fig:lower-gaussian} presents the simulation for different lengths $n$. As $n$ gets bigger, $M(t_{\theta_{\min}})$ converges to $-1$ with vanishing variance. Hence, we observe a tight lower bound $M_{\min}=-1$ for Gaussian distributions.

\section{Non-trivial Bounds for Other Distributions}

In this section, we present non-trivial memory strength constraints that we have found for several other distributions, including (a) Gamma distribution, (b) Student's t-distribution, (c) Weibull distribution, and (d) log-normal distribution. In Figure \ref{fig:other-distributions}, for each distribution, we plot simulated $M(t_{\theta_{\max}})$ and $ M(t_{\theta_{\min}}) $ against the  parameter that controls the shape of the distribution. Gamma, Weibull and log-normal distribution are also parameterized by a parameter that controls the scale ($\lambda$ for Gamma and Weibull, $\mu$ for log-normal), which does not affect the memory strength.

\end{document}